\documentclass{article}

\usepackage{graphicx}
\usepackage{amsmath}
\usepackage{amsfonts}

\def\abstract#1{\vskip 7mm
        \begin{center}{\large \textbf{Abstract}}\par \smallskip
                \begin{minipage}[c]{12cm}
                        \small #1
                \end{minipage}
        \end{center}
}
\def\title#1{\begin{center}{\Large\bf #1}\end{center}}
\def\author#1{\vskip 5mm \begin{center}{#1}\end{center}}
\def\address#1{\begin{center}{\it #1}\end{center}}
\makeatletter

\begin{document}

\title{%
Undeformed (additive) energy conservation law in Doubly Special
Relativity }
\author{%
\textbf{Gianluca Mandanici}\footnote{E-mail:
gianluca.mandanici@istruzione.it} }
\address{%
Dipart.~Fisica, Univ.~Roma ``La Sapienza'', P.le Moro 2, 00185 Roma, Italy}

\abstract{%
All the Doubly Special Relativity (DSR) models studied in
literature so far involve a deformation of the energy conservation
rule that forces us to release the hypothesis of the additivity of
the energy for composite systems. In view of the importance of the
issue for a consistent formulation of a DSR statistical mechanics
and a DSR thermodynamics, we show that DSR models preserving the
usual (i.e. additive) energy conservation rule can be found. These
models allow the construction of a DSR-covariant extensive energy.
The implications of the analysis for the dynamics of DSR-covariant
multiparticle systems are also briefly discussed.}

\section{Introduction}
Doubly Special Relativity has been proposed as a possible
two-scale extension of Special Relativity~
\cite{Amelino-Camelia:2000ge,Amelino-Camelia:2000mn,Kowalski-Glikman:2001gp,Bruno:2001mw,dsr2}.
The introduction of the second invariant scale $\lambda$,
eventually connected with the Planck length/energy, leads to some
departures from Special Relativity already manifesting at the
kinematical level. It has been clear from the
beginning~\cite{Amelino-Camelia:2000ge}, that the two most
immediate consequences of the introduction of this new invariant
scale are: for a single particle, the possibility to get a
modification of the energy-momentum dispersion relation (and a
consequent possible modification of the relation between
particle's velocity and particle's energy, see also~\cite{vdsr,
vncst}); for composite systems, the modification of the
energy-momentum conservation rules (see also~\cite{DSRP} and
references therein).

On a theoretical ground however, whereas the modification of the
free propagation does not seem to lead to inconsistencies, the
modification of the energy-momentum conservation laws appears to
be troublesome. In \cite{JW} (see also \cite{MS1}) a procedure to
construct all-order (in the Planck scale) energy-momentum
conservation rules has been proposed. When applied to the known
DSR models, this procedure produces modifications of both the
energy and the spatial momentum conservation rule. Similar
modifications are encountered adopting procedures based on
non-trivial co-products (see~\cite{LN,KL,GL} and references
therein).

In these scenarios, in which the energy and the spatial momentum
conservation laws are both modified, the deformation of the energy
conservation rule appears to be particularly troublesome since it
directly leads to difficulties in the construction of
DSR-covariant multiparticle systems, statistical mechanics and
thermodynamics. Nevertheless DSR-based statistical mechanics has
been proposed in \cite{KSM} and there are a number of studies in
the field of cosmology and black-hole physics (see for example
\cite{AM,KR,AAMY,NS,PGAMM}) that construct thermodynamical
quantities from DSR-motivated energy-momentum dispersion
relations. As first noticed in \cite{RS}, the fact that the total
energy of the system cannot be covariantly defined as the sum of
the energies of the particles composing the system
($E_{system}\neq E_1+E_2+...+E_n$), has the implication that the
energy fails to be a fully DRS-covariant extensive quantity.

Moreover, if we define with $\bigoplus$ the composition law of the
energies and with $E_{1},E_{2},E_{3}$ the energies of three
subsystems which we think the original system composed of, for all
the models proposed in literature so far, we get $E_{S}=
E_{1}\bigoplus E_{2} \bigoplus E_{3}\neq E_{1}\bigoplus ( E_{2}
\bigoplus E_{3})\neq (E_{1}\bigoplus E_{2} )\bigoplus E_{3}$,
which means that the energy of the system should depend on the
particular way we order the subsystems. The same conclusion holds
for the mass of the system (i.e. the zero-momentum energy) that
should change depending on the particular choice of the
decomposition.

Another problem related to the missing additivity of the energy is
the so called ``spectator problem''. This problem manifests in the
fact that the evolution of a single particle strongly depends on
how one considers the particle as a part of a larger system. We
will discuss in more detail this problem in the next sections.

All the above mentioned problems are automatically absent in every
relativity scheme in which the energy of a set of particles is
simply the sum of the energies of the single particles composing
the system, as it is the case in Galilean Relativity and in
Special Relativity. In the next sections we will show that also in
DSR there is no need to renounce to the hypothesis of the
additivity of the energy.

\section{DSR models preserving the additivity of the energy}
\subsection{The general procedure}

We start our analysis by requiring the covariance of the
undeformed energy conservation law in a $n$-to-$m$ particle
scattering process:
\begin{align}
\sum_{\alpha=1}^n E^{(\alpha)}=\sum_{\beta=1}^m E^{(\beta)},
\label{EC}
\end{align}
under the boost action obtained by commutators of the type
$\delta_k E=\xi [N_k,E]$, where
$[N_k,E]=\pi_k(\vec{p},\lambda)$\footnote{We assume, as usual,
that the Lorentz group $SO(3,1)$ is undeformed.}. One gets:
\begin{equation}
\sum_{\alpha=1}^n
\pi^{(\alpha)}_k(\vec{p},\lambda)=\sum_{\beta=1}^m
\pi^{(\beta)}_k(\vec{p},\lambda). \label{mmc}
\end{equation}

The covariance of (\ref{mmc}) then implies that


\begin{equation}
  \sum_{\alpha=1}^n\sum_{i=1}^3 [N_j,p^{(\alpha)}_i]
  \frac{\partial \pi^{(\alpha)}_k}{\partial p_i}=\sum_{\beta=1}^m\sum_{i=1}^3 [N_j,p^{(\beta)}_i]
  \frac{\partial \pi^{(\beta)}_k}{\partial p_i}.
\end{equation}

The above equations are simply solved if the commutators
$[N_j,p_i]$ satisfy the relations
\begin{equation}
\sum_{i=1}^3[N_j,p_i]\frac{\partial \pi_k}{\partial
p_i}=E\delta_{jk}.\label{LS}
\end{equation}

The linear system (\ref{LS}) can be easily solved with respect to
$[N_i,p_j]$ providing the action of the boosts on the spatial
momenta:
\begin{align}
[N_i,p_j]&= E \left({\frac{\partial \pi}{\partial
p}}\right)^{-1}_{ij},
\end{align}
where we have used the shorthand notation
$\left(\frac{\partial\pi}{\partial p}\right)
_{ik}=\frac{\partial\pi_k}{\partial p_i}$.

 From the action of the boosts on the energy and on the spatial momenta one immediately deduces the form of the dispersion
relation:
\begin{align}
E^2- \vec{\pi}^2(\vec{p},\lambda)=m^2.
\end{align}

\subsection{An explicit realization} To do an explicit example of a
DSR model allowing undeformed energy conservation law, let us
consider boost actions given by the commutators:
\begin{align}
[N_i,E]&= \frac{p_i}{(1-\lambda p)},\\
[N_i,p_j]&= E(1-\lambda p)\left(\delta_{ij}-\lambda \frac{p_i
p_j}{p}\right).
\end{align}

The energy-momentum dispersion relation then is given by
\begin{align}
 E=\sqrt{m^2+\frac{p^2}{(1-\lambda p)^2}}.
\end{align}

It is straightforward to verify that, when the spatial momentum
approaches the (invariant) Planck momentum, $p\rightarrow
\lambda^{-1}$, the energy diverges.

Covariant generalizations of the conservation rules are:
\begin{align}
 \sum_{\alpha=1}^n E^{(\alpha)} &=\sum_{\beta=1}^m E^{(\beta)} ,\\
 \sum_{\alpha=1}^n \frac{\vec{p}^{~(\alpha)}}{(1-\lambda p^{~(\alpha)})} &=\sum_{\beta=1}^{m}
 \frac{\vec{p}^{~(\beta)}}{(1-\lambda p^{~(\beta)})},
\end{align}
that preserve the additivity of the energy.

\section{Implications for the multiparticle dynamics}
A ``spectator problem'' manifests in DSR when one tries to define
the dynamics of DSR systems through the usual canonical/Heisenberg
evolution scheme. Technically the problem is due to the fact that
the time-evolution operator is bilinear in its two arguments,
whereas the Hamiltonian of the system, which is the generator of
the time evolution, depends nonlinearly on the Hamiltonians of the
subsystems.

To be more specific let us consider a $n$-particle system whose
Hamiltonian is
\begin{align}
H_{S}=H_{S}\left(H_{1},H_{2},...,H_{n} \right),
\end{align}
and a generical observable $O^{(k)}(p^{(k)})$ associated to the
$k$-th particle of the system.

We get for the evolution of the observable
\begin{align}
 [O^{(k)}(p^{(k)}),H_{S}]=[O^{(k)}(p^{(k)}),H_k]\frac{\partial H_{S}(p^{(1)},p^{(2)},...,p^{(n)})}{\partial
 H_{k}},
\end{align}
If the additivity of the energy is missing, the evolution of the
observable $O^{(k)}(p^{(k)})$ depends, through the term
$\frac{\partial H_{S}(p^{(1)},p^{(2)},...,p^{(n)})}{\partial
H_{k}}$,  on the state of the other particles composing the
system. In this case it would be difficult even to define the
notion of ``free particle" and strong (long-range) interactions
would be hard to avoid.

If instead we adopt additive energy composition law, being
\begin{align}
  H_{S}=H_{1}+H_{2}+...+H_{n},
\end{align}
and
\begin{align}
 \frac{\partial H_{S}(p^{(1)},p^{(2)},...,p^{(n)})}{\partial
 H_{k}}=1,
\end{align}
we obtain
\begin{align}
 [O^{(k)}(p^{(k)}),H_{S}]=[O^{(k)}(p^{(k)}),H_k],
\end{align}
which is free from the unlikely non-local effects.

\section{Comparison with a previous analysis}

In Ref.~\cite{DR} a model has been studied in which both the
energy and the spatial momentum compose linearly. The
energy-momentum dispersion relation analyzed in Ref.~\cite{DR} is
the one originally proposed in~\cite{MS1}:
\begin{align}
E^2=p^2+m^2\left(1+\frac{2k}{m}E\right).\label{formula}
\end{align}

The authors of Ref.~\cite{DR} find that the energy-momentum
dispersion relation (\ref{formula}) is covariant under the linear,
but inhomogenus, boost action given by
\begin{align}
p'^{\mu}=\Lambda^{\mu}_{\nu}p^{\nu}+(\delta^{\mu}_{0}-\Lambda^{\mu}_{0})m
k,\label{EC}
\end{align}
where the Einstein notation has been used for the energy-momentum
vector $p^{\mu}\equiv(E,\vec{p})$ and for the Lorentz
transformation tensor $\Lambda^{\mu}_{\nu}$. $k$ is a deformation
parameter.

The linearity of the composition of both the energy and the
spatial momentum follows directly from the linearity of the boost
action (\ref{EC}).

However, from the point of view of a scattering process, as
analyzed in the previous sections, important differences emerge
with respect to the special relativistic conservation laws. In
fact, if we consider an $n$-to-$m$ particle scattering process,
the request of covariance of the conservation laws:
\begin{align}
\sum_{\alpha=1}^n E^{(\alpha)}=\sum_{\beta=1}^m E^{(\beta)},
\\\sum_{\alpha=1}^n \vec{p}^{~(\alpha)}=\sum_{\beta=1}^m
\vec{p}^{~(\beta)},
\end{align}
 under the action of (\ref{EC}), also implies that
\begin{align}
 \sum_{\alpha=1}^n k m_{(\alpha)}=\sum_{\beta=1}^m k m_{(\beta)}.\label{constrain}
\end{align}

Equation (\ref{constrain}) results as a constraint on the allowed
particle scattering. To better understand the meaning of this
constraint we have to make explicit the relation among the
parameter $k$, the particle mass $m$, and the Planck parameter
$\lambda$ we used in this paper. Various possibilities have been
discussed in the same Ref.~\cite{DR}. The cases there taken in to
the account are: $k\propto\lambda$, $k\propto m\lambda$ and
$k\propto \lambda / m$.

In the first case ($k\propto\lambda$) we would get
\begin{align}
 \sum_{\alpha=1}^n m_{(\alpha)}=\sum_{\beta=1}^m  m_{(\beta)}, \label{rel1}
\end{align}
in the second case ($k \propto m\lambda$) we would get
\begin{align}
 \sum_{\alpha=1}^n m_{(\alpha)}^2=\sum_{\beta=1}^m  m_{(\beta)}^2,\label{rel2}
\end{align}
whereas, in the third case ($k\propto\lambda/m$), we would obtain
$n=m$.

In the first two cases ($k\propto\lambda$, $k\propto m \lambda$)
the constraints regard the relations between the masses of the
incoming particles and that of the outcoming particles. In the
third case ($k\propto\lambda /m$), the constraint only allows
scattering processes preserving the particle number ($n=m$). All
the constraints we have found are violated in the observed
scattering processes\footnote{Notice that for massless particles
the constraints are trivially satisfied but, in that case, the
model simply reduces to Special Relativity.}. Thus, we have to
conclude that the model presented in Ref.~\cite{DR} is not viable
from the perspective adopted in this paper.

\section{Conclusions}
We have shown that the non-additivity of the energy is not a
common feature of DSR theories. In the framework of DSR theories
with additive-energy composition law one avoids many problems that
mine the consistency of DSR statistical mechanics and DSR
thermodynamics. In particular: $i)$ the energy (and the mass) of a
composite system is a extensive quantity; $ii)$ the energy (and
the mass) of a system does not depend on how one thinks a system
decomposed in subsystems; $iii)$ the time evolution of
multiparticle systems, obtainable by means of the usual
canonical/Heisenberg formalism, is free from long-range nonlocal
effects.

\section*{Acknowledgements}

The author thanks Giovanni Amelino-Camelia for the valuable
suggestions, and Daniele Oriti for ospitality at DAMTP in
Cambridge (UK) where this work was initiated. The author also
thanks the authors of Ref.~\cite{DR} for having brought their
paper to his attention.

\end{document}